\documentclass[aps,prl,twocolumn,superscriptaddress]{revtex4}
\usepackage{amsmath}
\usepackage{empheq}
\usepackage{color}
\usepackage{ulem}
\begin{document}

\title{Non-equilibrium conductivity at quantum critical points}
\author{A. M. Berridge} 
\affiliation{London Centre for Nanotechnology, University College London, 17-19 Gordon St, London, WC1H 0AH, UK}
\author{A. G. Green}
\affiliation{London Centre for Nanotechnology, University College London, 17-19 Gordon St, London, WC1H 0AH, UK}
\date{\today}

\begin{abstract}
Quantum criticality provides an important route to revealing universal non-equilibrium behaviour. A canonical example of a quantum critical point is the Bose-Hubbard model, which we study under the application of an electric field. A Boltzmann transport formalism and $\epsilon$-expansion are used to obtain the non-equilibrium conductivity and current noise. This approach allows us to explicitly identify how a universal non-equilibrium steady state is maintained, by identifying the rate-limiting step in balancing Joule heating and dissipation to a heat bath. It also reveals that the non-equilibrium distribution function is very far from a thermal distribution.
\end{abstract}

\maketitle


\section{Introduction}

Even when driven far from equilibrium it is hoped that quantum systems will be governed by a set of general principles. While a few such principles have been identified, they are far less constraining than their equilibrium counterparts~\cite{Jarzynski,Crooks}. Quantum criticality provides a useful route to identify others. The universal temporal scaling of quantum critical systems is inherited by their non-equilibrium steady-state, providing an important class of universal results. This has been shown in a few cases~\cite{DalidovichPhillips,GreenSondhi,GreenVishwanath,KarchSondhi,SonnerGreen,KirchnerSi}, however the approaches used often assume a steady-state and circumvent the subtleties of the underlying physics that permit a universal steady-state to form. Here we explicitly show how a non-equilibrium steady-state is reached.

We study the Bose-Hubbard model as a canonical example of a quantum critical point. We show that the non-equilibrium distribution function is far from thermal, and that an expansion about a thermal distribution at some effective temperature is not sufficient to capture the out-of-equilibrium state. Instead we show that a distribution that is highly elongated along the field direction is a good variational solution of the Boltzmann equation. This is consistent with previous approaches based on $1/N$ expansions~\cite{GreenSondhi} or the AdS/CFT correspondance~\cite{KarchSondhi,SonnerGreen}, which also find highly non-equilibrium steady states. Within this picture the current noise can be calculated using a Boltzmann-Langevin approach. This procedure gives a Johnson noise form at some effective noise temperature, consistent with previous results ~\cite{GreenVishwanath,SonnerGreen}.

A central task is to show how the steady state is maintained by balancing heat flows in the system. The Landau-Zener or Schwinger mechanism produces particle-hole pairs from the vacuum~\cite{GreenSondhi}. Acceleration of these charges by the electric field produces heat through Joule heating. This heat is removed via a heat sink at the edge of the sample in order to reach a steady-state. In order to establish a universal out-of-equilibrium distribution, the rate limiting step for this process must be universal. This implies that the scattering of energy into the thermal transport modes must be slower than the transport of energy to the edge of the sample - since the latter will depend upon the size and shape of the sample. Achieving this requires that the Wiedemann-Franz law be broken - this is possible in a system with several species of charge carrier, because of the different components of the scattering integral involved in thermal and electrical scattering \footnote{Subtler issues are possibly at play in the breakdown of the Wiedemann-Franz law in metals. See M. A. Tanatar et al , Science 316,  1320 (2007)}. Previous analyses were able to side-step these considerations \footnote{$1/N$ expansion or probe-brane approximations in the AdS setting}. A benefit of our current approach is the we confront this issue head-on and make the conditions for a universal steady-state explicit.

The structure of the paper is as follows: We begin by considering the inhomogeneous Boltzmann equation and demonstrate how spatial gradients driving heat flow to the sample edge may be replaced by a universal sink term. We will show that attempts to solve this equation by expanding around a thermal distribution fail, so that the non-equilibrium distribution is far from thermal. We then show that a distribution in which particles and holes are strongly Lorentz boosted in opposite directions is a good variational ansatz for the Boltzmann equation. Using this ansatz we calculate the current noise through a Boltzmann-Langevin approach, ending with a discussion of our results and their connection to other work.

\textit{Boltzmann equation} --- We analyze the response of the Bose-Hubbard model at its particle-hole symmetric point to an applied electric field. This model describes bosons hopping on a lattice with an on-site interaction. It is a benchmark model of quantum criticality, with many tools developed to analyze its equilibrium behaviour. At the particle-hole symmetric point, its effective theory is essentially a Klein-Gordon theory with a $\phi^4$ interaction~\cite{FisherFisher} - we give an explicit form for this in the supplementary materials. This supports bosonic normal modes of positive and negative charge, which we will refer to as particles and holes. We calculate the distribution function describing the occupation of these normal modes in the non-equilibrium steady-state, therefore obtaining the non-equilibrim conductivity and current noise.

We use a Boltzmann transport approach~\cite{DamleSachdev}. An $\epsilon$-expansion is used to control scattering, calculating the scattering integral in 3-$\epsilon$ dimensions. Particle-hole pair production appears as a source term~\cite{GreenSondhi}. We show that heat flow to the bath can be represented by a spatially homogeneous sink term. This explicitly demonstrates the balancing of the various processes and the establishment of a universal steady state.

The quantum Boltzmann equation describes scattering processes between the normal modes and their response to external fields. The occupation of the modes is represented by a Wigner distribution function $f^\pm_{\bf k}({\bf x}, t)$ where +/- represent the positively/negatively charged modes (`particles'/`holes')~\footnote{The Wigner distribution is defined as $f_{\bf k}=\int d{\bf q} \langle a^\dagger_{\bf k+q}a_{\bf k-q}\rangle/2\pi^2$ where $a^\dagger (a)$ are creation(annihilation) operators for the normal modes and angular brackets represent averages over the state of the system.}. The equation describing the evolution of this distribution function is~\cite{DamleSachdev}:
\begin{eqnarray}
\left(\partial_t + {\bf v}\cdot\partial_{\bf r} \mp {\bf E}\cdot\partial_{\bf k}\right) f_{\bf k}^\pm
&=&
\mathcal{S}_{\bf k}[f^\pm_{\bf q}] + g_{\bf k}^{source}.
\label{Beq}
\end{eqnarray}
The left  of this equation describes the evolution of the distribution function from the bare dynamics of the particles. The time derivative term is neglected since we are considering steady-state solutions. We show that when the thermal conductivity is large, as appropriate for the Bose-Hubbard model, the gradient term ${\bf v}\cdot\partial_{\bf r}$ can be replaced with a homogeneous sink term describing scattering into thermal transport modes. The third term on the left describes the acceleration of particles by the electric field. On the right, the first term represents scattering between particles and holes~\cite{DamleSachdev}. The form of this term is discussed in more detail in the supplementary information. The second term is the source term representing particle-hole pair production from the vacuum via the Landau-Zener mechanism and may be derived by solving the equation of motion for the anomalous distribution~\cite{GreenSondhi},
\begin{eqnarray}
g^{source}_{\bf k}
&=&
\frac{\pi}{4} \sqrt{E} e^{-\pi k^2/E}.
\end{eqnarray}
The key step in solving this equation is to set up a spatially homogeneous Boltzmann equation for the non-equilibrium steady state. This can be done when the thermal conductivity is extremely large, leading to small thermal gradients.

\textit{Spatially homogeneous limit and universal response} --- The universal response is set by the rate-limiting step in dissipating heat. Joule heating and production of particle-hole pairs needs to be balanced by transport to a heat-bath at the sample edge. If transport to the edge is the rate limiting step, then thermal gradients build up across the sample and the response is non-universal, depending on the size and geometry of the sample. If, however, scattering into the heat-carrying modes is the rate limiting step, the gradients are small and the response is universal. In this case, the thermal gradients can be replaced by a homogeneous sink term that describes scattering into the heat carrying modes.

In the case of the Bose-Hubbard model the thermal conductivity is infinite as there are no processes that relax energy and momentum. Scattering into thermal transport modes is therefore the rate limiting step and the response is universal. The electrical conductivity, however, is still finite as the two species of charge carrier allow for current relaxation~\cite{DamleSachdev}.

In order to show this explicitly, we solve the Boltzmann equation including a temperature gradient across the sample. Consider an explicit example with a heat sink at the transverse boundary. We expand the distribution function in spatial gradients~\cite{TremblayPatton}, $f^\pm_{\bf k}(y) = f^\pm_{\bf k} + h^{(0)}_{\bf k} + y h^{(1)}_{\bf k} + y^2 h^{(2)}_{\bf k}$, where $y$ is the coordinate in a direction transverse to the flow of electrical current. The Boltzmann equation (\ref{Beq}) can then be solved in terms of the zero-modes of the linearised scattering integral (these are discussed further in the supplementary materials, and Refs.[\onlinecite{BhaseenGreenSondhi,MullerFritzSachdev,FritzSachdev}]). Such zero modes are due to conservation of energy, particle number and momentum and are therefore present whatever distribution we expand about. When the thermal conductivity is large the gradient terms $h^{(1)}_{\bf k}$ and $h^{(2)}_k$ become small - little variation in the distribution function is required to transport heat effectively. To leading order the spatially homogeneous part of the solution is given by $h^{(0)}_{\bf k} = -\mathcal{S}^{-1}\left(\sigma^z_{ph} {\bf E}\cdot\partial_{\bf k} f^\pm_{\bf k} + g^{source}_{\bf k} - \alpha v^y h_{\bf k}^\pm\right)$, where $v^y$ is the group velocity in the direction of heat transport, $h_{\bf k}^\pm$ is a zero-mode of the linearised scattering integral related to thermal transport and $\alpha$ is a constant which must be determined. The solution is reproduced by solving a homogeneous Boltzmann equation with an additional sink term $g^{sink}_{\bf k} = \alpha v^y h_{\bf k}^\pm$. This sink describes scattering into the heat-carrying modes, removing energy from the system. The presence of the (near) zero-mode~\footnote{In the infinite clean system this is exactly a zero-mode, we imagine a case where the thermal conductivity is extremely large, but not formally infinite.} reflects the fact that the thermal conductivity is large, manifesting as the inverse of the scattering integral being dominated by the zero-mode.

In this way we obtain a universal steady-state spatially homogeneous Boltzmann equation:
\begin{eqnarray}
\mp {\bf E}\cdot\partial_{\bf k} f^\pm_{\bf k} &=& \mathcal{S}_{\bf k}[f^\pm_{\bf q}] + g^{source}_{\bf k} - g^{sink}_{\bf k}.
\end{eqnarray}
Spatial gradients are replaced by a universal sink term representing scattering of particles into the heat-carrying modes. The physics of this response is encoded in the zero-modes of the scattering integral, the existence of which depends only upon the conservation laws of the system. The general form of the sink term is therefore universal. We have explicitly chosen to calculate the sink term for scattering into a mode that carries heat transversely across the sample. However, since the form of the steady-state distribution is  determined by integrals of the sink term, this choice does not affect the results. Having set up this homogeneous Boltzmann equation we now need to solve it for the distribution function. 

\textit{Failure of expansion about a thermal distribution} --- One might anticipate that the steady state distribution is close to a thermal distribution at some self-consistent effective temperature. However we show that an expansion about a thermal distribution fails and introduce an alternative ansatz based upon this insight.

\begin{figure}
\includegraphics[width=3in]{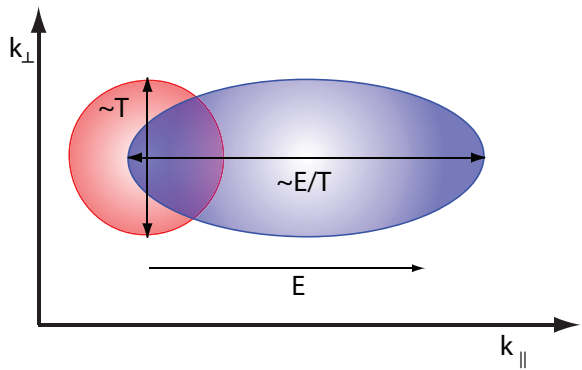}
\caption{\label{fig:Scaling}(Color online) Sketch of the generic argument for the distribution being far from thermal when the prefactor of the effective temperature is small. The momentum shift in the field direction is proportional to $E/T_{eff}$ while the energy of the distribution is proportional to $T_{eff}$. The parameter which controls the relative size of these is the effective temperature prefactor $T_{eff} = {\tilde T}_{eff} \sqrt{E}$, which is found to be small in the $\epsilon$-expansion.}
\end{figure}

We motivate this argument physically by considering the scaling properties of the Boltzmann equation, sketched in Fig.(\ref{fig:Scaling}). The energy of a distribution is set by the effective temperature, which we know from the scaling of the source term is proportional to the square root of the electric field, $T_{eff} = \sqrt{E} {\tilde T}_{eff}$~\cite{GreenSondhi}, where ${\tilde T}_{eff}$ is a universal dimensionless number. The effect of electric field is to introduce a momentum shift to the distribution. The size of this shift is given by $E/T_{eff} = \sqrt{E}/{\tilde T}_{eff}$. If the energy is greater than the momentum shift then the distribution is close to thermal and an expansion is appropriate. If the energy is less than the momentum shift the distribution is far from thermal and an expansion about a thermal distribution will fail. Which regime we are in is set by the numerical prefactor ${\tilde T}_{eff}$.

There are no tunable parameters in this argument. At the quantum critical point the validity of the expansion is set purely by ${\tilde T}_{eff}$. This parameter is ultimately determined by the detailed form of the scattering integral and must be found numerically. It turns out that the value obtained within an $\epsilon$-expansion is extremely small and therefore the expansion fails. Physically, this is because the effect of the electric field dominates and scattering procesess are not strong enough to relax the distribution back to nearly thermal.

We make this argument precise by performing the expansion about a thermal distribution and finding that it does not converge. The full procedure is detailed in the supplementary materials. The main steps are as follows: We expand the distribution function in particle-hole symmetric and anti-symmetric deviations from a thermal distribution, at an effective temperature set by number and energy conservation. Solving for these deviations order-by-order in the control parameter $\epsilon$, the first order term is the linear-response solution obtained by Damle and Sachdev~\cite{DamleSachdev}, albeit at a self-consistently determined effective temperature. However, in the present case each term in the series grows and the expansion does not converge. The non-equilibrium distribution is, therefore, far from thermal and cannot be accessed by an expansion about this state. We will take the opposite limit in which the distribution is highly distorted by the electric field.

\textit{The highly-boosted distribution} --- The lesson is that an electric field leads to a distribution that is very elongated along the field axis. Recognising this, we use a thermal distribution Lorentz boosted antisymmetrically for particles and holes as a variational ansatz:
\begin{eqnarray}
f^\pm_{\bf k} &=& f^T\left(\frac{k \pm {\bf v}\cdot{\bf k}}{T_{eff} \sqrt{1-v^2}}\right),
\label{eq:dist}
\end{eqnarray}
where $f^T(\epsilon_{\bf k}/T)$ denotes a thermal distribution and ${\bf v}$ is the boost velocity in the direction of electric field. The magnitude of ${\bf v}$ will be determined as a variational parameter. Since we expect the distribution to be highly elongated, $v$ will be close to one. We use this limit to simplify the scattering integral and show that the calculated value of $v$ is self-consistent.

The sink term for the boosted distribution is given by the zero-modes of the scattering integral linearised about an antisymmetric, boosted distribution. It is precisely this mode that would carry a heat current to the boundary of the sample. For a heat current in the transverse $y$-direction $h_{\bf k} = k_y \partial_k f^\pm_{\bf k}$ and $g_{\bf k}^{sink} = \alpha k_y^2 \partial_k f_{\bf k}^\pm$.

In order to complete the solution we must find the undetermined parameters, $\alpha$ - the sink term prefactor, ${\tilde T}_{eff}$ - the effective temperature prefactor, and $v$ - the boost velocity. These are found by taking three moments of the Boltzmann equation. Two of these moments represent number and energy conservation, and show the balance between source, sink and Joule heating. Summing over particle species, then integrating over all ${\bf k}$, gives an equation representing number conservation:
\begin{eqnarray}
0 &=& \int d{\bf k} \, g_{\bf k}^{source} - \int d{\bf k} \, g_{\bf k}^{sink},
\label{eq:mom1}
\end{eqnarray}
where we clearly see the role of the sink in removing particles from the system. We note that in the $\epsilon$-expansion all integrals are carried out in three spatial dimensions, with the value of $\epsilon$ then setting the dimensionality of the final result. Multiplying by the dispersion $\epsilon_{\bf k}$ before integrating gives the equation for energy conservation:
\begin{eqnarray}
\sigma_E |E|^2 &=& 2 \int d{\bf k} \, \epsilon_{\bf k} g_{\bf k}^{source} - 2 \int d{\bf k} \, \epsilon_{\bf k} g_{\bf k}^{sink}.
\label{eq:mom2}
\end{eqnarray}
where $\sigma_E$ is the electrical conductivity and the left hand side of the equation represents Joule heating. In both of these equations the scattering integral integrates to zero since it conserves number and energy. We need a final moment to fix all three parameters, which we obtain by multiplying by $\epsilon_{\bf k}^2$ before summing over ${\bf k}$:
\begin{eqnarray}
\lefteqn{\sum_{\sigma =\pm 1} \int d{\bf k} \, \epsilon_{\bf k}^2 \sigma  {\bf E}\cdot\partial_{\bf k} f^\pm_{\bf k}}
\nonumber\\ &&
=  2 \int d{\bf k} \, \epsilon_{\bf k}^2 \mathcal{S}_{\bf k}[f^\pm_{\bf q}]  + 2 \int d{\bf k} \, \epsilon_{\bf k}^2 g_{\bf k}^{source} - 2 \int d{\bf k} \, \epsilon_{\bf k}^2 g_{\bf k}^{sink}.
\nonumber\\
\label{eq:mom3}
\end{eqnarray}
The dependence of the solution upon scattering enters through this equation. In the limit of a distribution of the form (\ref{eq:dist}) with $v \sim 1$ its evaluation simplifies dramatically. The phase space for in-scattering is very restricted compared to that of out-scattering and we find $\mathcal{S}_{\bf k} [f_{\bf k}^\pm] = \Gamma_{\bf k} f_{\bf k}^\pm$ with $\Gamma_{\bf k} = -2\pi^3 \epsilon^2/75 \ T_{eff}^2/k$.

This approximation of dominant out-scattering is very similar to that used in the $1/N$ expansion. In that case it came from an ingenious use of the expansion, in which the electric field coupled to only one of $N$ modes. Out-scattering into $N-1$ modes dominates over in-scattering into one particular mode. Here the approximation is justified self-consistently by the emergent dynamics.

The three equations (\ref{eq:mom1},\ref{eq:mom2},\ref{eq:mom3}) allow us to solve for $\alpha$, ${\tilde T}_{eff}$ and $v$. They may be solved numerically, but as our approximation to the scattering integral is valid in the high-boost limit, an approximation with $v$ close to one is sufficient. In this limit,  the coefficients may be determined analytically with the result that $\alpha \simeq 8.14$, ${\tilde T}_{eff} \simeq 0.2$ and $v \simeq 0.96$, which is self-consistent with our approximations.

The conductivity in two dimensions  is given by:
\begin{eqnarray}
\sigma_E =\frac{{\bf j}\cdot{\bf E}}{E^2}
&=&
\sum_\sigma \sigma \int d{\bf k} \frac{{\bf E}\cdot{\bf k}}{E^2 k} f^\sigma_{\bf k}
=
\frac{\pi}{6 \sqrt{2 \delta v}}{\tilde T}_{eff}^2,
\end{eqnarray}
where $\delta v = 1-v$. This evaluates to $\sigma_E=0.074 \ q^2/\hbar$, where we have introduced the factor $q^2/\hbar$, previously set to one, with $q$ the charge of the particles. We compare this to the equilibrium result obtained by Damle and Sachdev of $\sigma_E = 0.165 \ q^2/\hbar$. Comparing with the $1/N$ results~\cite{GreenSondhi,KrempaSachdev} we see that there the conductivity is also reduced in the non-equilibrium case, although by a smaller amount.

\begin{figure}
\includegraphics[width=3in]{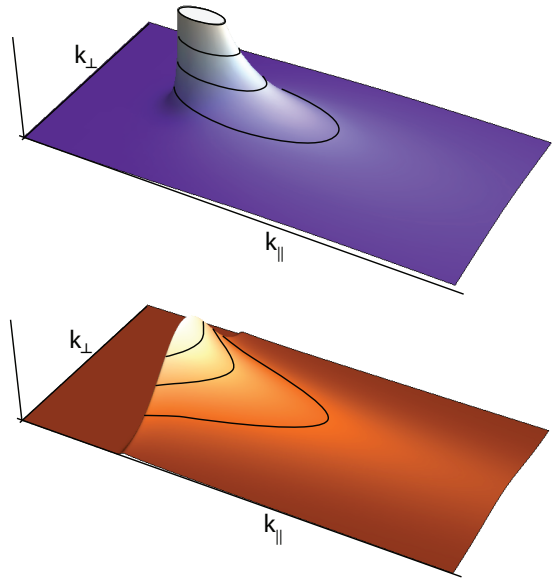}
\caption{\label{fig:Comparison}(Color online) Top: An illustration of the highly antisymmetrically boosted distribution Eq.(\ref{eq:dist}). Bottom: An illustration of the distribution found in the $1/N$ method.}
\end{figure}

\textit{Current noise} --- Having found the universal non-equilibrium distribution function, we calculate the current fluctuations. Our starting point is the Boltzmann-Langevin equation describing fluctuations of the distribution function $\Delta f_{\bf k}$ \cite{Kogan1969,Tremblay1982}.
\begin{eqnarray}
\left( \partial_t + {\bf v}\cdot\partial_{\bf r} + {\bf E}\cdot\partial_{\bf k}\right) \Delta f_{\bf k}
&=&
\mathcal{S}_{\bf k,q} \Delta f_{\bf k} + \eta_{\bf k},
\end{eqnarray}
where $\eta$ is a noise term. This approach assumes that scattering processes are independent and so Poisson distributed. The variance of the noise is equal to the mean scattering rate. The correlations of $\eta_{\bf k}$ are given by~\cite{GreenVishwanath}:
\begin{eqnarray}
\langle \eta_{\bf k}({\bf r},t) \eta_{{\bf k}'}({\bf r}',t')\rangle 
&=&
(2\pi)^2 \delta_{{\bf r}, {\bf r}'} \delta_{t,t'} \delta({\bf q}-{\bf q}') \Gamma_q f_{\bf q}
\nonumber\\
\end{eqnarray}
In the limit of long times and large distances the gradient terms of the Boltzmann-Langevin equation can be ignored. To first order we ignore the momentum derivative also and obtain the fluctuation in occupation number as $\Delta f_{\bf k} = \eta_{\bf k}/\Gamma_k$. The current fluctuations are therefore~\cite{GreenVishwanath}:
\begin{eqnarray}
\langle j_\alpha({\bf r},t) j_\beta({\bf r}',t')\rangle 
&=&
2 e^2 \delta_{{\bf r}, {\bf r}'} \delta_{t,t'} \delta_{\alpha,\beta} \int d{\bf k} \frac{f_{\bf k}}{\Gamma_q}
\nonumber\\
&\simeq&
\delta_{{\bf r}, {\bf r}'} \delta_{t,t'} \delta_{\alpha,\beta} \frac{50 e^2 \sqrt{E} {\tilde T}}{\pi \epsilon^2}\frac{1}{2 \delta v}.
\nonumber\\
\end{eqnarray}
If we choose to identify this result with Johnson noise at an effective temperature $T_{eff}'$, we find that the appropriate temperature is not the same as that associated with the steady state distribution. Writing the current noise in the form $4 \sigma { T}_{eff}' \sqrt{E}$ gives an effective noise temperature ${T}_{eff}' \simeq 2.3 \sqrt{E}$. This is of the same order as- though larger than - the temperature of a thermal distribution that has the same energy per mode as the out-of-equilibrium distribution \footnote{The energy per mode in the distribution Eq.(\ref{eq:dist}) is $\frac{\pi^4}{30 \zeta(3)} \frac{4}{3\sqrt{2\delta v}} T_{eff}$, whereas the energy per mode in a thermal distribution at temperatue $T$ is $\frac{\pi^4}{30 \zeta(3)} T$. Note that the temperature ${\tilde T}'_{eff}$ entering the current noise is not the same as the parameter ${\tilde T}_{eff}$ characterising the steady state distribution}. Interestingly, an effective noise temperature higher than the energy scale characterising the steady state was previously found for a thin metallic system equilibrated by phonons~\cite{Tremblay1982}. 

\textit{Discussion} --- We have shown both that the non-equilibrium steady state can be explicitly set-up by balancing heat flows, and that the resulting distribution is very far from thermal, being extended along the direction of the applied field. 

That a steady-state is established had been assumed in previous work, and implemented by exploiting the $1/N$ approach, and in holography where the probe brane limit plays a similar role. In the $\epsilon$-expansion we must explicitly include the processes by which heat is transported out of the sample. We have shown that this is a universal process as follows: The thermal conductivity of the model is extremely large, and vanishingly small thermal gradients are sufficient to drive a compensating heat flow. Scattering into thermal transport modes is rate-limiting, and this scattering is universal. The effect of heat flow to the boundary may therefore be encoded in a sink term added to the Boltzmann equation.

The distribution function is highly elongated along the field direction and far from a thermal distribution. It cannot therefore be accessed by an expansion about a thermal distribution at some effective temperature. Strongly Lorentz boosting the charge carriers in opposite directions provides a good variational ansatz. We have shown self-consistently that the boost velocity in such a distribution is close to one. Such a distribution is consistent with the highly asymmetric distribution found in the $1/N$ case, as illustrated in Fig.(\ref{fig:Comparison}). Further evidence of the profound non-equilibrium nature of the steady state is provided by the different temperature scales characterising the current noise and the steady-state distribution. Curiously, other calculations have found that both the response function and fluctuations depend upon the same effective temperature and are related by an apparent fluctuation dissipation relation~\cite{KirchnerSi, SonnerGreen} at all frequencies even though the steady state itself is characterised by a different effective temperature.

We end with some words about the feasibility of realising these results experimentally. This may be possible either in cold atomic gasses - in which the Bose Hubbard model has been realised~\cite{GreinerBloch} along with effective fields through potential gradients or otherwise~\cite{LinSpeilman} - or else in solid state systems with particle-hole symmetry such as graphene. In both cases, the main limitation is the difficulty of appropriate coupling to a bath. This may be accompanied by thermal boundary resistivity~\cite{Tremblay1982}. In addition, any deviation from particle-hole symmetry may lead to significant constraints upon the size of system over which universal results may be observed~\cite{Hogan2008}. One fascinating way around this is suggested in an elegant recent work, Ref.\cite{BernardDoyon} where the system in effect acts as its own thermal bath with the universal non-equilibrium steady state in a limited region of the sample.

\textit{Acknowledgements} --- We thank Joe Bhaseen for useful discussions. This research was supported by the EPSRC under grant code EP/I004831/1.

\end{document}